\documentclass[sigconf,9pt, nonacm]{acmart}

\usepackage{algorithm}
\usepackage{algorithmic}
\usepackage{subfigure}
\usepackage{amsmath}
\usepackage{mathtools}
\usepackage{multirow}
\usepackage{xcolor}
\usepackage{comment}
\usepackage{color}
\usepackage{bm}
\usepackage{etoolbox}
\usepackage{url}

\AtBeginDocument{%
  \providecommand\BibTeX{{%
    \normalfont B\kern-0.5em{\scshape i\kern-0.25em b}\kern-0.8em\TeX}}}

\newcommand\scaleup{0.6}
\newcommand\scale{0.38}

\AtBeginDocument{%
  \providecommand\BibTeX{{%
    \normalfont B\kern-0.5em{\scshape i\kern-0.25em b}\kern-0.8em\TeX}}}

\acmYear{2021}\copyrightyear{2021}
\setcopyright{acmcopyright}
\acmConference[IoTDI '21]{International Conference on Internet-of-Things Design and Implementation}{May 18--21, 2021}{Charlottesvle, VA, USA}
\acmBooktitle{International Conference on Internet-of-Things Design and Implementation (IoTDI '21), May 18--21, 2021, Charlottesvle, VA, USA}
\acmPrice{15.00}
\acmDOI{10.1145/3450268.3453519}
\acmISBN{978-1-4503-8354-7/21/05}

\begin{document}

\title{DeepObfuscator: Obfuscating Intermediate Representations with Privacy-Preserving Adversarial Learning on Smartphones}

\author{Ang Li$^1$, Jiayi Guo$^2$, Huanrui Yang$^1$, Flora D. Salim$^3$, Yiran Chen$^1$}
\affiliation{%
  \institution{$^1$Department of Electrical and Computer Engineering, Duke University\\
  $^2$Department of Automation, Tsinghua University\\
  $^3$Computer Science and Information Technology, School of Science, RMIT University}
  $^1$\{ang.li630, huanrui.yang, yiran.chen\}@duke.edu, $^2$guo-jy20@mails.tsinghua.edu.cn, $^3$flora.salim@rmit.edu.au
}






\begin{abstract}
Deep learning has been widely applied in many computer vision applications, with remarkable success. However, running deep learning models on mobile devices is generally challenging due to the limitation of computing resources. A popular alternative is to use cloud services to run deep learning models to process raw data. This, however, imposes privacy risks. Some prior arts proposed sending the features extracted from  raw data (e.g., images) to the cloud. Unfortunately, these extracted features can still be exploited by attackers to recover  raw images and to infer embedded private attributes (e.g., age, gender, etc.). In this paper, we propose an adversarial training framework, \textit{DeepObfuscator}, which prevents the usage of the features for reconstruction of the raw images and inference of private attributes. This is done while retaining useful information for the intended cloud service (i.e., image classification). DeepObfuscator includes a learnable encoder, namely, obfuscator that is designed to hide privacy-related sensitive information from the features by performing our proposed adversarial training algorithm. The proposed algorithm is designed by simulating the game between an attacker who makes efforts to reconstruct raw image and infer private attributes from the extracted features and a defender who aims to protect user privacy. By deploying the trained obfuscator on the smartphone, features can be locally extracted and then sent to the cloud.  
Our experiments on CelebA and LFW datasets show that the quality of the reconstructed images from the obfuscated features of the raw image is dramatically decreased from 0.9458 to 0.3175 in terms of multi-scale structural similarity (MS-SSIM). The person in the reconstructed image, hence, becomes hardly to be re-identified. The classification accuracy of the inferred private attributes that can  be achieved by  the attacker is significantly reduced to a random-guessing level, e.g., the accuracy of gender is reduced from 97.36\% to 58.85\%. As a comparison, the accuracy of the intended classification tasks performed via the cloud service is only reduced by 2\%. We also demonstrate the efficiency of DeepObfuscator, showcasing real-time performance of the deployed models on smartphones.
\end{abstract}


\maketitle

\section{Introduction}\label{sec:intro}
In the past decade, deep learning has achieved great success in many computer vision applications, such as face recognition \cite{parkhi2015deep} and image segmentation \cite{liu2015semantic}. However, running deep learning models on mobile devices is technically challenging due to limited computing resources. Many large-sized deep-learning-based applications are often deployed on cloud servers, i.e., ML-as-a-service (MLaaS), such as Amazon Rekognition, Microsoft Cognitive Services, etc. These cloud-based services require users to send data (e.g., images) to the cloud service provider. However, this requirement may raise users' concerns about privacy leakage, since various private information may be contained in the images (e.g., age, gender, etc.). One widely adopted solution to address this privacy issue is to upload only the extracted features rather than the raw image \cite{osia2017privacy,osia2018deep}. Unfortunately, the extracted features still contain rich information which can breach users' privacy. Specifically, an attacker can exploit the eavesdropped features to reconstruct the raw  image, and hence the identity of the person on the raw image can be discovered from the reconstructed image \cite{mahendran2015understanding}. In addition, the extracted features can also be exploited by an attacker to infer private attributes, such as gender, age, etc. 
Such adversary models can be trained by an attacker through continuously querying the cloud service to collect the eavesdropped features as inputs, and the ground truth of the queried data can be used as the labels.
\begin{figure*}[ht]
    \centering
    \includegraphics[scale=0.4]{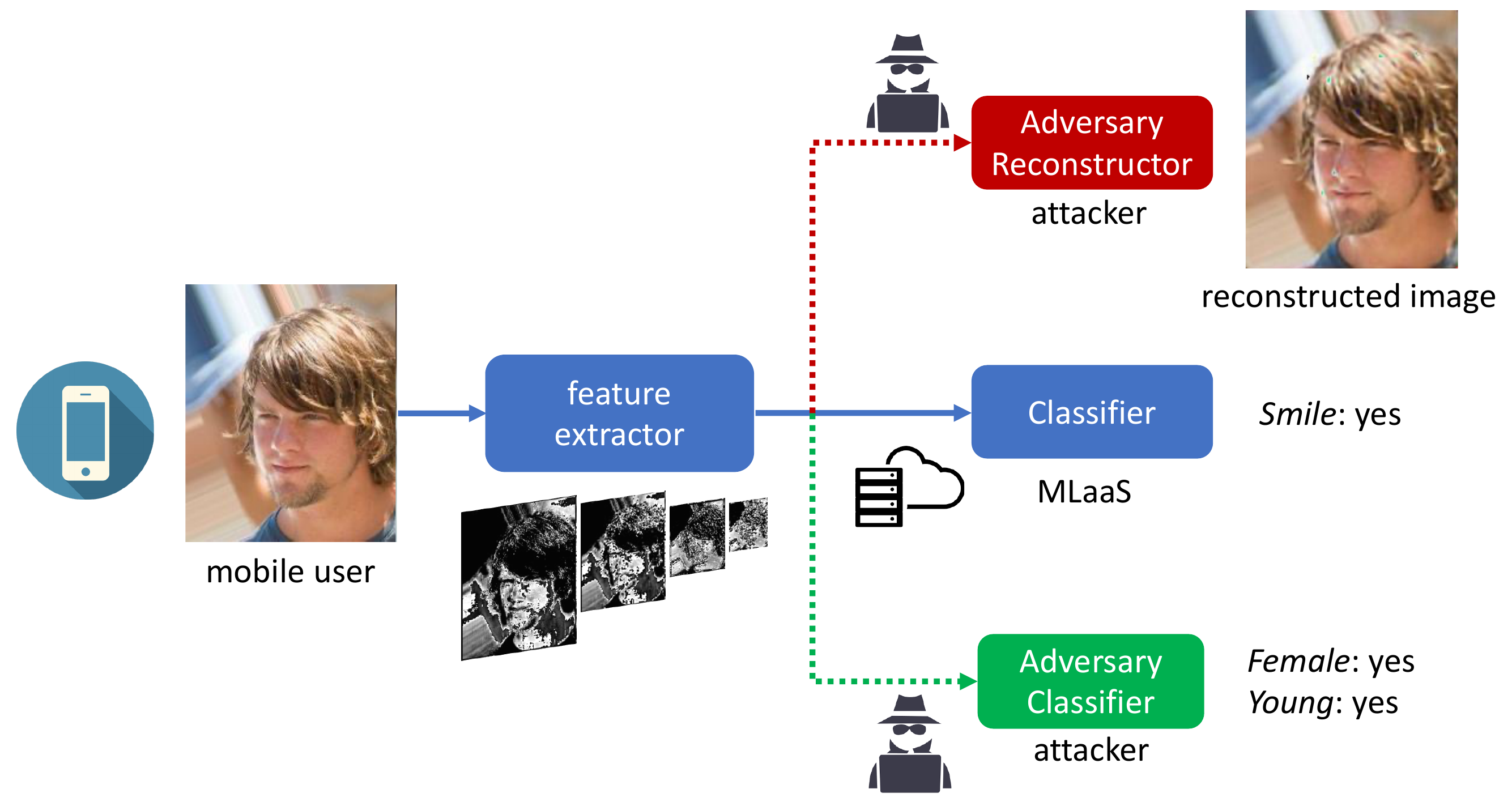}
    \caption{An example of reconstruction attack and private attribute leakage in a cloud service for facial attribute recognition.}
    \label{fig:example}
\end{figure*}

Figure \ref{fig:example} shows an example where the reconstruction attack and private attribute leakage occur in a MLaaS for facial attribute recognition. When a mobile user uploads an image for detecting facial attribute, the encoder will extract features and then send the features to the cloud server. The classifier deployed on the server takes the extracted features as inputs and then predicts whether the person in the received image is smiling or not. Note that a complete model is jointly trained in an end-to-end manner, and then split into the encoder and the classifier. However, the extracted features can be eavesdropped by an attacker. By continuously querying the cloud service, the attacker can collect the eavesdropped features to train an decoder, which is denoted as the adversary reconstructor, for recovering the raw  images. Besides, the eavesdropped features can also be exploited to train an adversary classifier for inferring the private attributes associated with the raw images.

There are a few studies have been performed to defend against reconstruction attacks. They perturbed either the raw data \cite{he2017differential} or the extracted features \cite{osia2017privacy,osia2018deep} through adding random noises. But these methods inevitably incur accuracy drop. Feutry \textit{et al.} \cite{feutry2018learning} proposed an image anonymization approach to hide sensitive features related to private attributes. However, none of these previous studies has investigated whether defending against only reconstruction attack or private attribute leakage is sufficient to prevent either or both types of attacks.

\begin{figure}[ht]
    \centering
    \subfigure{\includegraphics[scale=\scale]{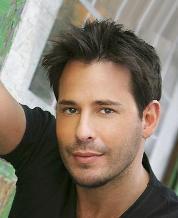}}\addtocounter{subfigure}{-1}
    \subfigure{\includegraphics[scale=\scale]{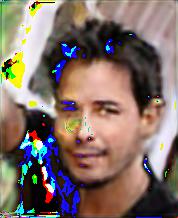}}\addtocounter{subfigure}{-1}
    \subfigure{\includegraphics[scale=\scale]{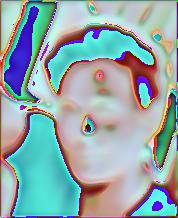}}\addtocounter{subfigure}{-1}\\
    \subfigure[]{\includegraphics[scale=\scale]{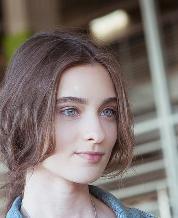}}
    \subfigure[]{\includegraphics[scale=\scale]{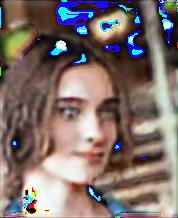}}
    \subfigure[]{\includegraphics[scale=\scale]{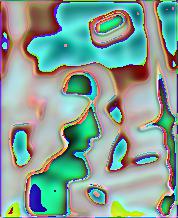}}
    \caption{Reconstructed images: defending against only private attribute leakage vs.  defending against only reconstruction attack. Column (a) is raw images, column (b) shows the reconstructed images when we  defend against only private attribute leakage, and column (c) displays reconstructed images when defending against only reconstruction attack.}
    \label{fig:motivation}
\end{figure}
 
Our experiments demonstrate that defending only reconstruction attack cannot prevent private attribute leakage, and vice versa. For example, Figure \ref{fig:motivation} shows two examples of reconstructed images when we defend against only either the reconstruction attack or private attribute leakage. The column (a) shows the raw images, and the column (b) displays the reconstructed images when we defend only private attribute leakage. These reconstructed images still contain many details of the raw images, and allow an attacker to re-identify the person in the images. The reconstructed images when defending against only reconstruction attack are shown in the column (c). Although almost all distinguishable information has been masked, we can still achieve a 93.7\% accuracy in detecting the gender of the person in the reconstructed images.
More details about this experiment can be found in Section \ref{subsec:motivation}. 

\begin{figure}[]
    \centering
    \includegraphics[scale=0.25]{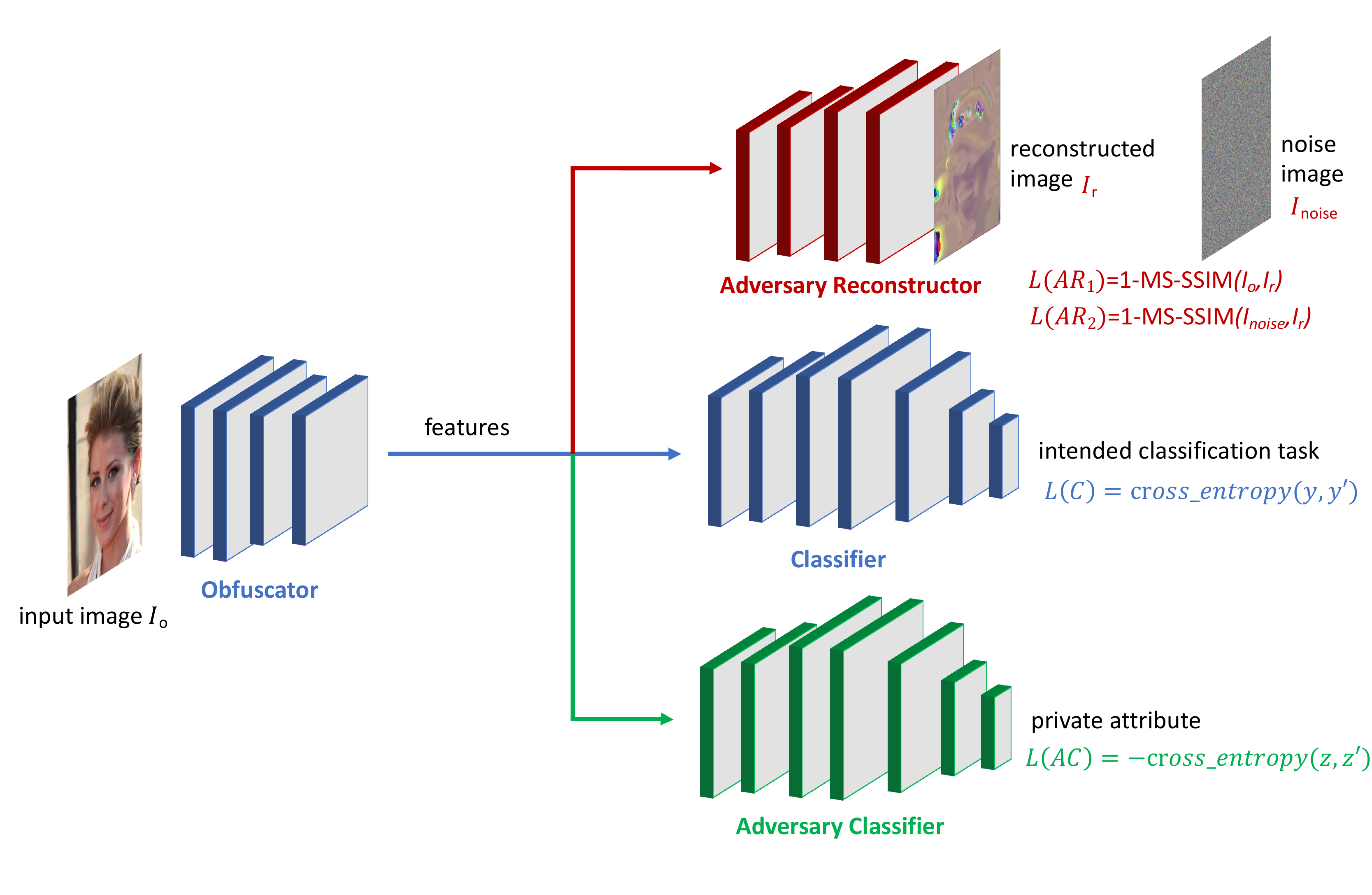}
    \caption{The design of DeepObfuscator.}
    \label{fig:design}
\end{figure}

In this work, we propose DeepObfuscator -- an adversarial training framework to learn an obfuscator that can hide sensitive information that can be exploited for reconstructing raw images and inferring private attributes, and still keep useful features for image classifications. Although we focus on image classifications in this paper, DeepObfuscator can be easily extended to many other tasks, e.g., speech recognition. By deploying the trained obfuscator on the smartphone, features can be locally extracted and then sent to MLaaS provider. As Figure \ref{fig:design} shows, DeepObfuscator consists of four modules: \textit{obfuscator}, \textit{classifier}, \textit{adversary reconstructor} and \textit{adversary classifier}. The key idea is to apply adversarial training for maximizing the reconstruction error of the adversary reconstructor and the classification error of the adversary classifier, but minimizing the classification error of the intended classifier. 

The main contributions of this paper are summarized as follows:
\begin{enumerate}
    \item We design DeepObfuscator which is an adversarial training framework that can simultaneously defend against both reconstruction attack and private attribute leakage while maintaining the accuracy of primary learning tasks;
    \item We are the first to experimentally demonstrate that defending against only the reconstruction attack or private attribute leakage is \textit{not} inclusive to each other;
    \item We quantitatively evaluate DeepObfuscator on CelebA and LFW datasets. The results show that the quality of reconstructed images from the obfuscated features is significantly decreased from 0.9458 to 0.3175 in terms of MS-SSIM, indicating that the person on the image is hardly 
    reidentifiable visually. The classification accuracy of the inferred private attributes is reduced by around 30\% to a random-guessing accuracy, but the accuracy of the intended classification tasks performed via the cloud service is reduced by only 2\%.
    \item We demonstrate the efficiency of DeepObfuscator, showcasing real-time performance of the deployed models on smartphones.
\end{enumerate}

The rest of this paper is organized as follows: Section \ref{sec:related} reviews the related work. Section \ref{sec:design}  elaborates the design of DeepObfuscator. Section \ref{sec:evaluation} evaluates the performance of DeepObfuscator. Section \ref{sec:case} shows two case studies of applying DeepObfuscator. Section \ref{sec:discuss} discusses the limitation of our proposed method. Section \ref{sec:conclusion} concludes this work.

\section{Related Work}\label{sec:related}
A large number of works have been done to protect data privacy using various anonymization techniques including \textit{k}-anonymity \cite{sweeney2002k}, \textit{l}-diversity \cite{mahendran2015understanding} and \textit{t}-closeness \cite{li2007t}. 
However, these solutions are designed for protecting sensitive attributes in a static database, and hence are not suitable to our addressed problem -- obfuscating intermediate representations of data while retaining the utility for DNN inference. 
Differential privacy \cite{duchi2013local,erlingsson2014rappor,bassily2015local,qin2016heavy,smith2017interaction,avent2017blender,wang2017locally} is another widely applied technique to prevent an individual's data record from being leaked with a strong theoretical guarantee. But the privacy guarantee provided by differential privacy is different from the privacy protection offered by DeepObfuscator. The goal of differential privacy is to inject random noise to a user’s  data record such that an adversary cannot identify the existence of this data record in the database. 
Different from differential privacy, our goal is to hide private information from the intermediate representations, such that an adversary cannot accurately infer the protected private information and successfully reconstruct the raw data. 
Li \textit{et al.}~\cite{li2020tiprdc} present an information theoretic approach to hide private information in features while maximally retaining the information carried by the raw data, such that the extracted features still have high utility for training DNN models. However, defending against the reconstruction attack is not taken into account by this method.
In addition, encryption-based approaches \cite{gilad2016cryptonets,yonetani2017privacy} have been presented to protect data privacy, but they require to train specialized DNN models on the encrypted data. Unfortunately, such encryption-based solutions prevent general dataset release and introduce substantial computational overhead.

Osia \textit{et al.} \cite{osia2020hybrid} combine dimensionality reduction, noise injection and Siamese fine-tuning to protect sensitive information from features, but it does not consider defending against the reconstruction attack.
De-identification is another popular  privacy-preserving method to prevent the identity from being visually recognized in many computer vision applications. 
There are various techniques to achieve de-identification, such as Gaussian blur \cite{oh2016faceless}, identity obfuscation \cite{oh2016faceless}, mean shift filtering \cite{winkler2014trusteye} and adversarial image perturbation \cite{oh2017adversarial}. 
Although those approaches are effective in protecting visual privacy, they all degrade the utility of the data for DNN inference. 

With recent developments of deep learning, several works have been proposed to protect data privacy by exploiting adversarial learning.
Pittaluga \textit{et al.} \cite{pittaluga2019learning} propose an adversarial learning method to learn  an encoder, aiming to defend against performing inference for specific attributes from the encoded intermediate representations. 
Seong \textit{et al.} \cite{oh2017adversarial} design an adversarial network to transform the raw image so that the attacker cannot successfully perform image recognition. 
Wu \textit{et al.} \cite{wu2018towards} present an adversarial framework to explicitly learn a degradation transform for the original video inputs in order to balance target task accuracy and the associated privacy budgets on the transformed video. 
Malekzadeh \textit{et al.} \cite{mobilesensor} propose an on-device transformation of sensor data that will be used for specific applications, such as monitoring selected daily activities. This method can prevent the user from being identified by the adversary. 
Liu \textit{et al.} propose PAN~\cite{liu2019privacy}, which is an adversarial learning framework to obfuscate features for the privacy purpose. However, PAN defenses against the reconstruction attack by enlarging the difference between the raw image and the reconstructed image in terms of pixel-wise distance, which may not be able to guarantee a significant perceptual difference between the raw image and the reconstructed image (see Section \ref{subsec:case}). In addition, PAN's performance is evaluated only for protecting single private attribute, which is not always the case in practice.

\section{Design of DeepObfuscator}\label{sec:design}
As Figure \ref{fig:design} shows, DeepObfuscator consists of three additional neural network modules: \textit{classifier} ($C$), \textit{adversary reconstructor} ($AR$) and  \textit{adversary classifier} ($AC$). The classifier works for the intended classification service. The adversary reconstructor and adversary classifier simulate an attacker in the adversarial training procedure, aiming to recover raw images and infer private attributes from the eavesdropped features. All the four modules are end-to-end trained using our proposed adversarial training algorithm.

Before presenting the details of each module, we give the following notations. We denote $D=\{I_1,\dots,I_N\}$ as the images in the dataset, where $N$ is the number of images, and $D'=\{I'_1,\dots,I'_N\}$ represents the reconstructed images that are generated by the adversary reconstructor. Let  $\mathcal{Y}=\{Y_1,\dots,Y_M\}$ denote the set of the target classes that the classifier is trained to predict, and $Y_i=\{y_{i1},\dots,y_{iN}\}$ denotes the corresponding labels of each class. Similarly, we adopt $\mathcal{Z}=\{Z_1,\dots,Z_K\}$ to denote the set of private classes that the adversary classifier aims to infer, and $Z_i=\{z_{i1},\dots,z_{iN}\}$ denotes the corresponding labels of each private class.

\subsection{Obfuscator}
The obfuscator ($O$) is a typical encoder which consists of an input layer, multiple convolutional layers, max-pooling layers and batch-normalization layers. The obfuscator is trained to hide privacy-related information while retaining useful information for intended classification tasks.

\subsection{Classifier}
The classifier ($C$) is jointly trained with the obfuscator as a complete CNN model. A service provider can choose any neural network architecture for the classifier based on task requirements and available computing resources. In DeepObfuscator, without loss of generality, we adopt a popular CNN architecture VGG16 \cite{simonyan2014very}, and split it into the obfuscator and the classifier.

The performance of the classifier $C$ is measured using the cross-entropy loss function, which is expressed as: 

\begin{equation}
   \mathcal{L}(C)=-\sum_{j=1}^{\mathcal{N}}\sum_{i=1}^{\mathcal{M}}y_{ij}log(y'_{ij})+(1-y_{ij})log(1-y'_{ij}), \label{eq:loss_c}
\end{equation}
where ($y_{1j},\dots,y_{Mj}$) denote the ground truth labels for the $j$th data sample, and ($y'_{1j},\dots,y'_{Mj}$) are the corresponding predictions. Therefore, the obfuscator and the classifier can be optimized by minimizing the above loss function as:
\begin{equation}
    \theta_o,\theta_c=\underset{\theta_o,\theta_c}{\arg\min} \mathcal{L}(C),
\end{equation}
where $\theta_o$ and $\theta_c$ are the parameters of the obfuscator and classifier, respectively.

\subsection{Adversary Classifier}
By continuously querying the cloud service, an attacker can train the adversary classifier ($AC$) using the eavesdropped features as inputs and the interested private attributes as labels. An attacker can infer private attributes via feeding the eavesdropped features to the trained adversary classifier. In DeepObfuscator, we apply the same architecture to both the classifier and the adversary classifier. However, the attacker can choose any architecture for the adversary classifier. As we shall show in Section \ref{subsec:eval_visible}, the performance of using different architectures in both the classifier and the adversary classifier will not be significantly different from the one that is achieved using the same architecture.

Similar to the classifier, the performance of the adversary classifier $AC$ is also measured using the cross-entropy loss function as: 
\begin{equation}
   \mathcal{L}(AC)=-\sum_{j=1}^{\mathcal{N}}\sum_{i=1}^{\mathcal{K}}z_{ij}log(z'_{ij})+(1-z_{ij})log(1-z'_{ij}), \label{eq:loss_ac}
\end{equation}
where ($z_{1j},\dots,z_{Mj}$) denote the ground truth labels for the $j$th eavesdropped feature, and ($z'_{1j},\dots,z'_{Mj}$) stand for the corresponding predictions. When we simulate an attacker who tries to enhance the accuracy of the adversary classifier as high as possible, the adversary classifier needs to be optimized by minimizing the above loss function as:
\begin{equation}
    \theta_{ac}=\underset{\theta_{ac}}{\arg\min} \mathcal{L}(AC),
\end{equation}
where $\theta_{ac}$ is the parameter set of the adversary classifier. On the contrary, when defending against private attribute leakage, we train the obfuscator in our proposed adversarial training procedure that aims to degrade the performance of the adversary classifier while improving the accuracy of the classifier. Consequently, the obfuscator can be trained using Eq. \ref{eq:adv_ac} when simulating a defender: 
\begin{equation}
    \theta_{o}=\underset{\theta_{o}}{\arg\min} \mathcal{L}(C)-\lambda_1\mathcal{L}(AC), \label{eq:adv_ac}
\end{equation}
where $\lambda_1$ is a tradeoff parameter. 

\subsection{Adversary Reconstructor} \label{subsec:adv_reconstructor}
The adversary reconstructor ($AR$), which is trained to recover the raw image from the eavesdropped features, also plays an attacker role. The attacker can apply any neural network architecture in the adversary reconstructor design. However, the worst case happens when an attacker knows the architecture of the obfuscator, and then builds the most powerful reconstructor, i.e., an exactly mirrored obfuscator by performing a layer-to-layer reversion. In DeepObfuscator, we adopt the most powerful reconstructor as the adversary reconstructor. The experiments in Section \ref{subsec:eval_visible} show our trained obfuscator can successfully defend against the brute-force reconstruction attack when an attacker trains the reconstructor with different neural network architectures.

When playing as an attacker, the adversary reconstructor is trained to optimize the quality of the reconstructed image $I_r$ as close as the original image $I_o$. In DeepObfuscator, we leverage MS-SSIM \cite{wang2003multiscale,ma2016group} to evaluate the performance of the adversary reconstructor, which is expressed as:
\begin{equation}
    \mathcal{L}(AR_1)=1-\operatorname{MS-SSIM}(I_o, I_r).\label{eq:loss_ar}
\end{equation}

The MS-SSIM value ranges between 0 and 1. The higher the MS-SSIM value is, the more perceptual similarity can be found between the two compared images, indicating a better quality of the reconstructed images. Consequently, an attacker can optimize the adversary reconstructor as:
\begin{equation}
    \theta_{ar}=\underset{\theta_{ar}}{\arg\min} \mathcal{L}(AR_1), \label{eq:loss_ar1}
\end{equation}
where $\theta_{ar}$ is the parameter set of the adversary reconstructor. On the contrary, a defender expects to degrade the quality of the reconstructed image as much as possible. To this end, we generate one additional Gaussian noise image $I_{noise}$. The adversary reconstructor is trained to make each reconstructed image similar to $I_{noise}$ but different from $I_o$, and the performance of the classifier should be maintained. When playing as a defender, the obfuscator can be trained as:
\begin{equation}
     \mathcal{L}(AR_2)=1-\operatorname{MS-SSIM}(I_{noise}, I_r)\label{eq:loss_ar2}
\end{equation}
\begin{equation}
     \theta_{o}=\underset{\theta_{o}}{\arg\min} \mathcal{L}(C)+ \lambda_2(\mathcal{L}(AR_2)-\mathcal{L}(AR_1)),\label{eq:adv_ar}
\end{equation}
where $\lambda_2$ is a tradeoff parameter.

\subsection{Adversarial Training Algorithm}\label{subsec:algorithm}
Algorithm \ref{alg:adv} summarizes the proposed four-stage adversarial training algorithm. Before performing the adversarial training, we first jointly train the obfuscator and the classifier without privacy concern to obtain the optimal performance on the intended classification tasks. Similarly, we also pre-train the adversary classifier and adversary reconstructor for initialization. As Algorithm \ref{alg:adv} shows, within each epoch of training, each adversarial training iteration consists of four batches. In the first two batches, we train the obfuscator to defend against the adversary reconstructor and the adversary classifier while keeping the classifier unchanged. For the third batch, we optimize the adversary reconstructor and the adversary classifier by simulating an attacker, but the parameters of the obfuscator and the classifier are fixed. Finally, we optimize the classifier to improve the classification accuracy on the intended tasks.

\section{Evaluation}\label{sec:evaluation}
In this section, we evaluate DeepObfuscator's performance on two real-world datasets, with a focus on the utility-privacy tradeoff. We also compare DeepObfuscator with existing solutions proposed in the literature and visualize the results. 

\subsection{Experiment Setup}\label{sec:exp-setup}
We implement DeepObfuscator with PyTorch, and train it on a server with 4$\times$NVIDIA TITAN RTX GPUs. We apply mini-batch technique in training with a batch size of 64, and adopt the AdamOptimizer \cite{adam} with an adaptive learning rate in all four stages in the adversarial training procedure. The architecture configurations of each module are presented in Table \ref{tb:model_arch}. We deploy the trained obfuscator on Google Pixel 2 and Pixel 3 to evaluate the real-time performance.

We adopt CelebA \cite{liu2015faceattributes} and LFW \cite{kumar2009attribute} for the training and testing of DeepObfuscator. CelebA consists of more than 200K face images. Each face image is labeled with 40 binary facial attributes. The dataset is split into 160K images for training and 40K images for testing. LFW consists of more than 13K face images, and each face image is labeled with 16 binary facial attributes. We split LFW into 10K images for training and 3K images for testing.

\begin{algorithm}[t]
\caption{Adversarial Training Algorithm}
\begin{algorithmic}[1]
\REQUIRE Dataset $\mathcal{D}$\\ 
\ENSURE $\theta_o, \theta_c, \theta_{ar}, \theta_{ac}$ \\
\STATE \textbf{Input:} Dataset $\mathcal{D}$
\FOR{$every\ epoch$}
\FOR{$every\ four\ batches$}

\IF{$batch\ idx\ mod\ 4 == 0$}
\STATE {\textbf{Defend against AR:}}\ 
\STATE {$ \mathcal{L}(C) + \mathcal{L}(AR_2) - \mathcal{L}(AR_1)\rightarrow \text{update}\ O(\theta_o)$}

\ELSIF{$batch\ idx\ mod\ 4 == 1$}
\STATE {\textbf{Defend against AC:}}\
\STATE {$ \mathcal{L}(C) - \mathcal{L}(AC)\rightarrow \text{update}\ O(\theta_o)$}

\ELSIF{$batch\ idx\ mod\ 4 == 2$}
\STATE {\textbf{reconstruction attack:}}
\STATE {$\mathcal{L}(AR_1)\rightarrow \text{update}\ AR(\theta_{ar})$}\ 
\STATE {\textbf{Infer private attributes:}}
\STATE {$\mathcal{L}(AC)\rightarrow \text{update}\ AC(\theta_{ac})$}\ 

\ELSE
\STATE {\textbf{Recover C:}}
\STATE {$\mathcal{L}(C)\rightarrow \text{update}\ C(\theta_c)$}\ 
\ENDIF
\ENDFOR
\ENDFOR
\end{algorithmic}\label{alg:adv}
\end{algorithm}

\begin{table}[t]
\centering
\caption{The architecture configurations of each module.}
\resizebox{0.48\textwidth}{!}{
\begin{tabular}{c|c|c}
\hline
\textbf{Obfuscator} & \textbf{Adversary Reconstructor} & \textbf{Classifier \& Adversary Classifier}\\
\hline

conv3-64& Upsample& 3$\times$conv3-256\\
\cline{2-2}
\cline{3-3}
conv3-64& deconv3-128& maxpool\\
\cline{1-1}
\cline{3-3}
maxpool& deconv3-64& 3$\times$conv3-512\\ 
\hline
conv3-128& Upsample& maxpool\\
\cline{2-3}
conv3-128& deconv3-64& 3$\times$conv3-512\\
\cline{1-1}
\cline{3-3}
maxpool& deconv3-3& maxpool\\
\cline{1-3}
&&2$\times$FC-4096\\
&&FC-label length\\
\cline{3-3}
&&sigmoid\\

\hline
\end{tabular}}\label{tb:model_arch}
\end{table}

\subsection{Motivation}\label{subsec:motivation}

Before presenting our performance evaluations, we first verify our motivation that defending against only reconstruction attack or private attribute leakage is not inclusive to each other. We apply our proposed adversarial training algorithm to defend against only one of these two attacks each time, and evaluate the attack performance on the other. In this experiment, we select `gender' as the private attribute. The results presented in Figure \ref{fig:motivation} verified our motivation. 

One na\"{\i}ve solution of the exclusion of defending against reconstruction attack and private attribute leakage is to first train an obfuscator to defend against one of these two scenarios using the adversarial training approach, and then continue to train the obfuscator to defend against the other one. However, this na\"{\i}ve solution can not simultaneously defend against both the scenarios because the parameters of the obfuscator keep being updated in the second step. The above limitation motivates the design of DeepObfuscator.

\subsection{Comparison Baselines}
We select four types of data privacy-preserving baselines \cite{liu2019privacy}, which have been widely applied in the literature, and compare them with DeepObfuscator. The details settings of the baseline solutions are presented as below.
\begin{itemize}
    \item \textbf{Noisy} method perturbs the raw data $x$ by adding Gaussian noise $\mathcal{N}(0,\sigma^2)$, where $\sigma$ is set to 40 according to \cite{liu2019privacy}. The noisy data $\Bar{x}$ will be delivered to the data collector. The Gaussian noise injected to the raw data can provide strong guarantees of differential privacy using less local noise. This scheme has been widely applied in federated learning \cite{papernot2018scalable,truex2019hybrid}. 
    \item \textbf{DP} approach injects Laplace noise the raw data $x$ with diverse privacy budgets \{0.1, 0.2, 0.5, 0.9\}, which is a typical differential privacy method. The noisy data $\Bar{x}$ will be submitted to the data collector. 
    \item \textbf{Encoder} learns the latent representation of the raw data $x$ using a DNN-based encoder. The extracted features $z$ will be uploaded to the data collector.
    \item \textbf{Hybrid} method further perturbs the above encoded features by performing principle components analysis (PCA) and adding Laplace noise \cite{osia2020hybrid} with varying noise factors privacy budgets \{0.1, 0.2, 0.5, 0.9\}.
\end{itemize}

\subsection{Effectiveness of Defending Against Reconstruction Attack}\label{subsec:eval_visible}
We quantitatively evaluate the quality of reconstructed images and obtain the results through a human perceptual study. Before showing quantitative results, we perform an experiment to simulate reconstruction attack using different reconstructor architectures. As introduced in Section \ref{subsec:adv_reconstructor}, we adopt the most powerful decoder, i.e., exactly the reverse of the obfuscator, as the adversary reconstructor for training. However, an attacker may not be able to know the architecture of the obfuscator, and hence the attacker may conduct a brute-force attack using the reconstructor with different architectures. We implement three additional reconstructors as attackers in our experiments. The architectural configurations of those reconstructors are presented in Table \ref{tb:attacker_model_arch}. URec\#1 and URec\#2 are built based on the architecture of U-net \cite{ronneberger2015u}, and ResRec is implemented with ResNet \cite{he2016deep} architecture. Each reconstructor is separately trained with the same pre-trained obfuscator. 

We again adopt the MS-SSIM to evaluate the quality of the reconstructed images that are generated by each reconstructor, i.e., comparing the similarity between the reconstructed image and the corresponding raw image. A smaller value of MS-SSIM implies less similarity between the reconstructed image and the raw image, indicating a more effective defense against reconstruction attacks. Table \ref{tab:avg_ssim} presents the average MS-SSIM for attacking reconstructors on testing data. The results show that although we apply the mirrored obfuscator architecture for the adversary reconstructor when training the obfuscator, the trained obfuscator can effectively defend against reconstruction attacks no matter what kinds of architecture are adopted by an attacker in the reconstructor design.

\begin{table}[t]
\centering
\newcommand{\tabincell}[2]{\begin{tabular}{@{}#1@{}}#2\end{tabular}}
\caption{Adversary reconstructor configurations.}
\begin{tabular}{c|c|c}
\hline
\textbf{URec\#1} & \textbf{URec\#2} & \textbf{ResRec}\\
\hline
\multicolumn{3}{c}{Input (54$\times$44$\times$128 feature maps)}\\
\hline
\tabincell{l}{conv3-64 \\ conv3-64 \\ conv3-64}&\tabincell{l}{conv3-64 \\ conv3-64}&\tabincell{l}{transconv3-64 \\ 
$
 \left[
 \begin{matrix}
   3\times3, & 64 \\
   3\times3, & 64
  \end{matrix}
  \right] \times 2
$}\\
\hline
\multicolumn{3}{c}{Upsample}\\
\hline
\tabincell{l}{conv3-128 \\ conv3-128 \\ conv3-128}&\tabincell{l}{conv3-128 \\ conv3-128}&\tabincell{l}{$
 \left[
 \begin{matrix}
   3\times3, & 64 \\
   3\times3, & 64
  \end{matrix}
  \right] \times 2
$}\\
\hline
\multicolumn{3}{c}{Upsample}\\
\hline
\tabincell{l}{conv3-256 \\ conv3-256 \\ conv3-256}&\tabincell{l}{conv3-256 \\ conv3-256}&conv1-3\\
\hline
\tabincell{l}{conv3-3 \\ conv1-3}&\tabincell{l}{conv3-3 \\ conv1-3}&sigmoid\\
\hline
sigmoid&sigmoid&\\

\hline
\end{tabular}\label{tb:attacker_model_arch}
\end{table}

\begin{table}[ht]
    \centering
        \caption{MS-SSIM for different attack reconstructors.}
    \resizebox{0.48\textwidth}{!}{
    \begin{tabular}{c|c|c|c|c}
    \hline
        \multirow{2}{*}{\textbf{Training Reconstructor}} &  \multicolumn{4}{|c}{\textbf{Attack Reconstructor}}\\
        \cline{2-5}
         & AR in DeepObfuscator & URec\#1 & URec\#2 & ResRec \\
        \hline
        AR in DeepObfuscator & 0.3175 & 0.3123 & 0.3095 & 0.3169 \\
        \hline
    \end{tabular}}

    \label{tab:avg_ssim}
\end{table}{}

\begin{figure}[ht]
    \centering
    \includegraphics[scale=0.45]{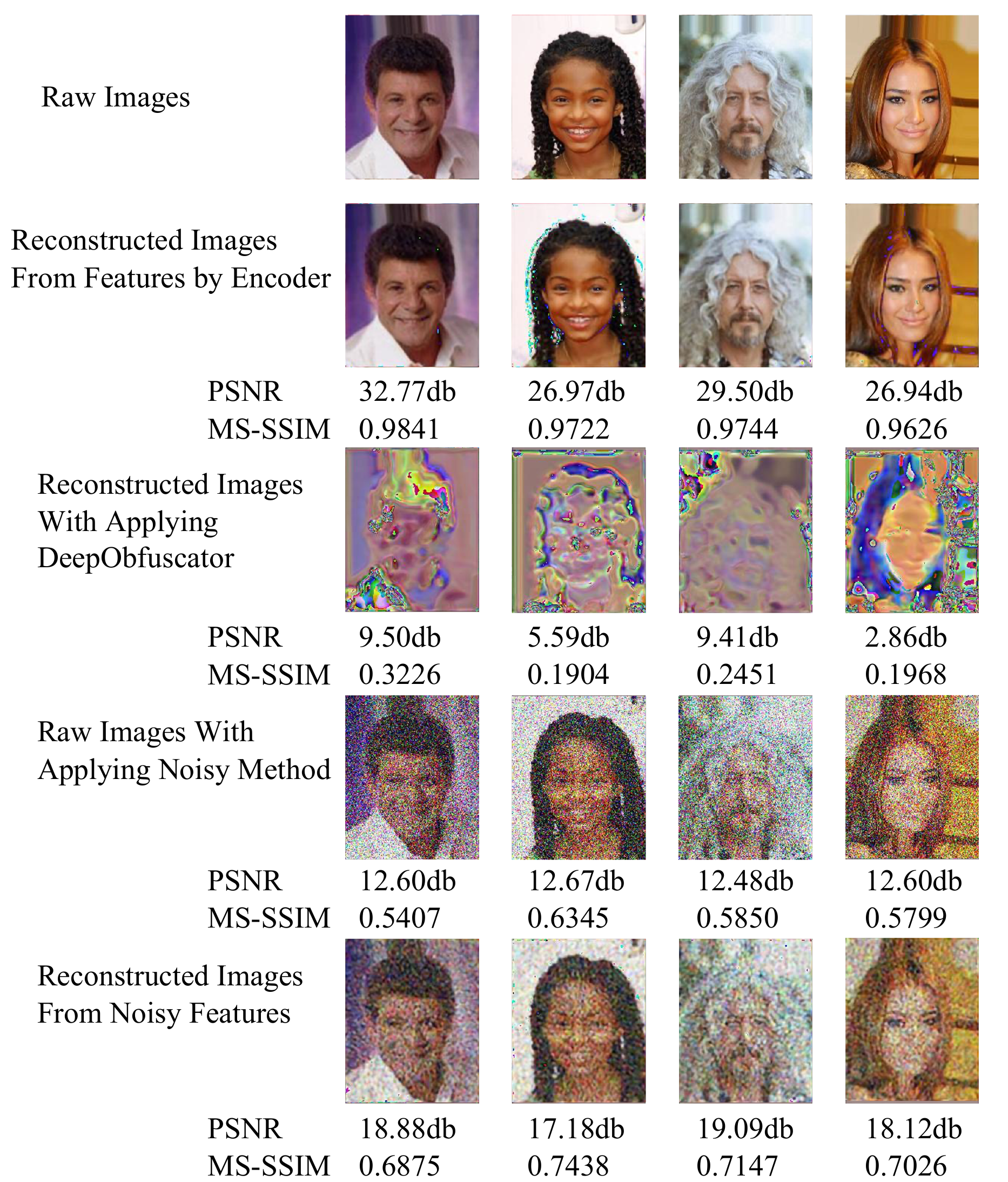}
    \caption{The comparison between reconstructed images with DeepObfuscator and two baseline models.}
    \label{fig:comparision}
\end{figure}

\textbf{Quantitative Evaluation.}
In addition to MS-SSIM, we also adopt the Peak Signal to Noise Ratio (PSNR) -- a widely used metric of image quality, to evaluate the quality of the reconstructed images.
In this experiment, the obfuscator is trained by setting the intended classification task as `smile' and the private attribute as `gender' in CelebA. Smaller values of MS-SSIM and PSNR indicate a stronger defense against reconstruction attack. Table \ref{tab:avg_psnr_ssim} presents the average MS-SSIM and PSNR on the testing data of DeepObfuscator and two baseline models. The result shows that DeepObfuscator is the most effective one to defend against reconstruction attack and the baseline models can hardly hide privacy information from the features. Figure \ref{fig:comparision} illustrates several examples of the reconstructed images. With our proposed adversarial training, the images that are reconstructed from the obfuscated features become unrecognizable. Even though directly applying Noisy method can hide more private information than apply Encoder method, the person in the image that is reconstructed from noisy features can still be re-identified. In summary, both quantitative evaluations and visual results show that DeepObfuscator can effectively defend against reconstruction attack.

\begin{table}[ht]
    \centering
    \caption{Average PSNR and MS-SSIM for DeepObfuscator and two baseline models.}
    \begin{tabular}{c|c|c|c}
    \hline
      \textbf{Metric}   &  \textbf{DeepObfuscator} & \textbf{Encoder} & \textbf{Noisy} \\
      \hline
      MS-SSIM   & 0.3175 & 0.9458 & 0.7263 \\
      \hline
      PSNR & 6.32 & 27.81 & 16.97 \\
      \hline
    \end{tabular}
    \label{tab:avg_psnr_ssim}
\end{table}

\begin{figure}[t]
    \centering
    \subfigure[]{\includegraphics[scale=0.4]{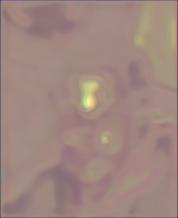}}
    \subfigure[ ]{\includegraphics[scale=0.4]{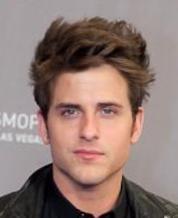}}
    \subfigure[]{\includegraphics[scale=0.4]{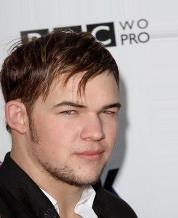}}
    \subfigure[]{\includegraphics[scale=0.4]{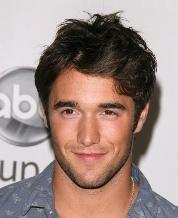}}
    \subfigure[]{\includegraphics[scale=0.4]{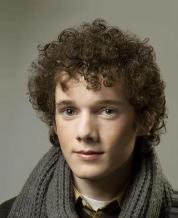}}
    \caption{An example question of the human perceptual study. (a) is the reconstructed image, and (b)-(e) are the four options.}
    \label{fig:survey}
\end{figure}

\textbf{Human Perceptual Study.}
We also conduct an online human perceptual study to directly examine whether a person in a reconstructed image can be re-identified by humans. This study consists of 10 questions, each of which includes one reconstructed image and four raw images as options. One of the four options contains the person in the reconstructed image. Participants are instructed to choose the option that looks like the person the most in the reconstructed image. There is no time limit about how long the participants can see these images.
Figure \ref{fig:survey} shows one example question in the survey. It is very difficult to find hints from the reconstructed image to identify the correct answer, i.e., Figure \ref{fig:survey}(e). There are 40 participants involved in this survey and each participant can only submit one response, hence, we collect 40 responses in total in this study. The average re-identification accuracy of 10 questions is 28\%, which is very close to a random guess, i.e., 25\% for 4 options. Furthermore, we can imagine if there is no option offered to an attacker, it will become more challenging to re-identify the person from the reconstructed image alone.

\begin{figure*}[t]
    \centering
    \subfigure[]{\includegraphics[scale=\scaleup]{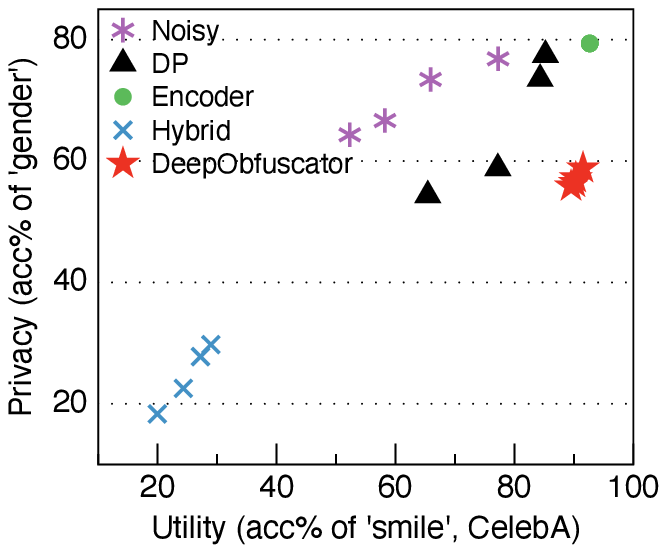}}
    \subfigure[]{\includegraphics[scale=\scaleup]{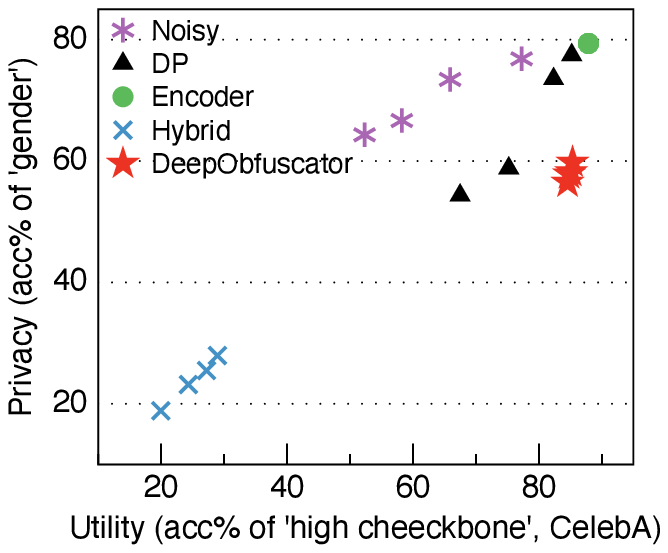}}
    \subfigure[]{\includegraphics[scale=\scaleup]{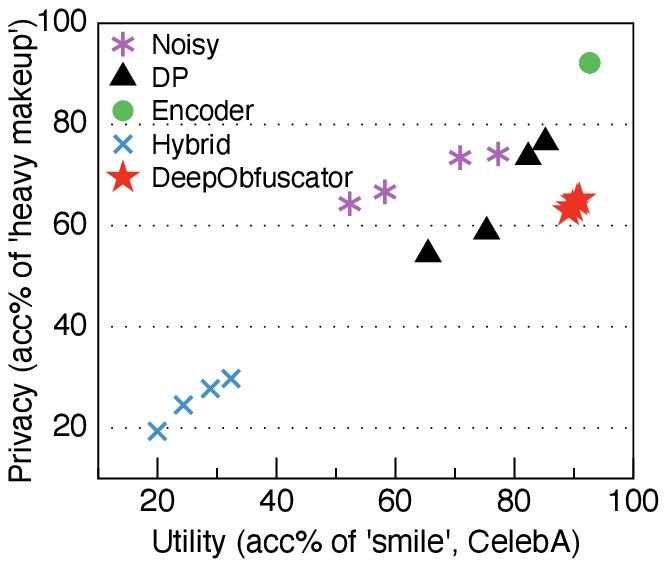}}
    \subfigure[]{\includegraphics[scale=\scaleup]{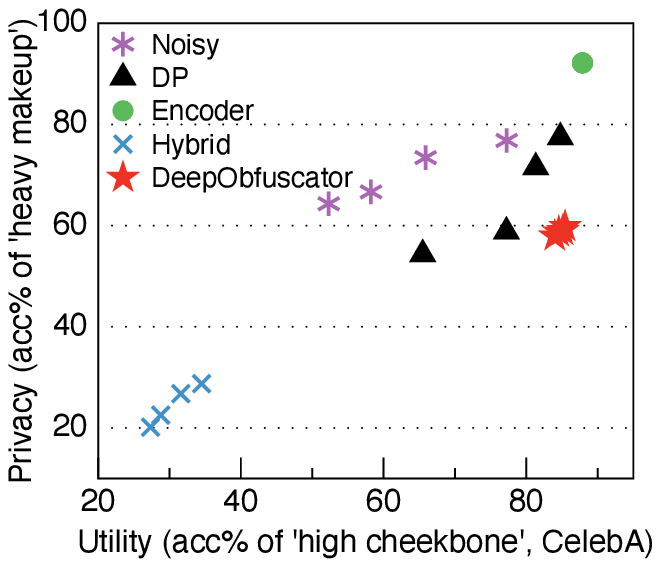}}
    \subfigure[]{\includegraphics[scale=\scaleup]{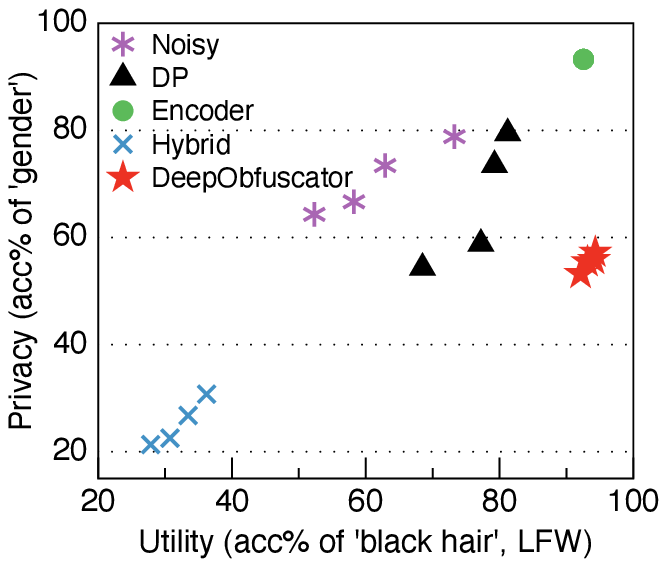}}
    \subfigure[]{\includegraphics[scale=\scaleup]{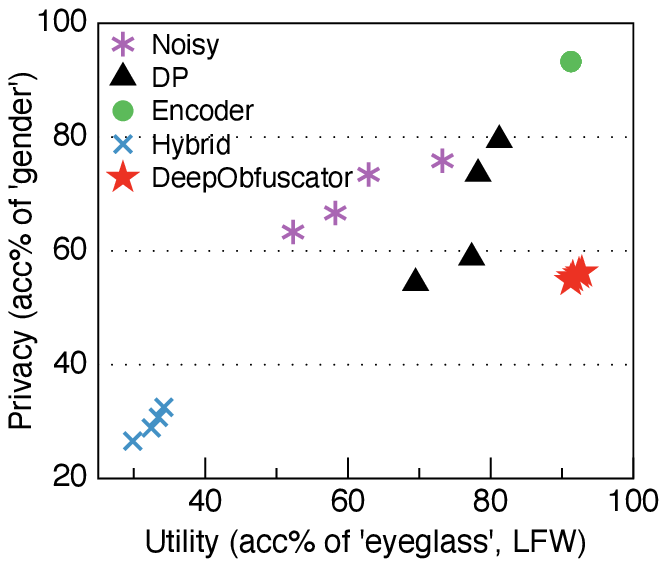}}
    \subfigure[]{\includegraphics[scale=\scaleup]{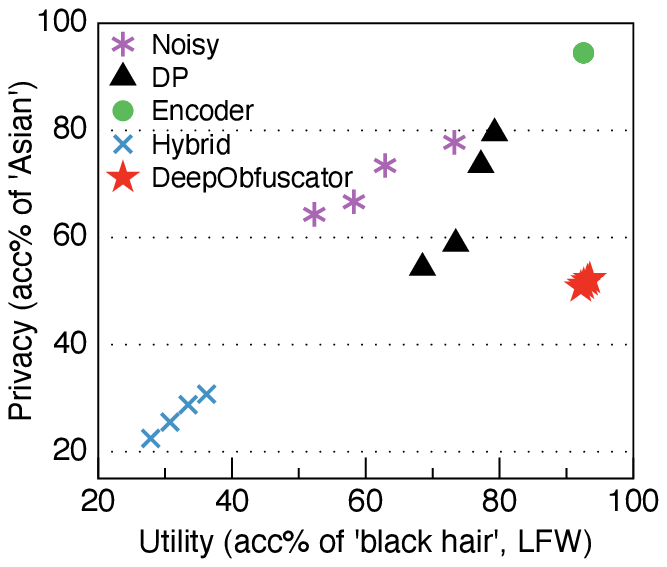}}
    \subfigure[]{\includegraphics[scale=\scaleup]{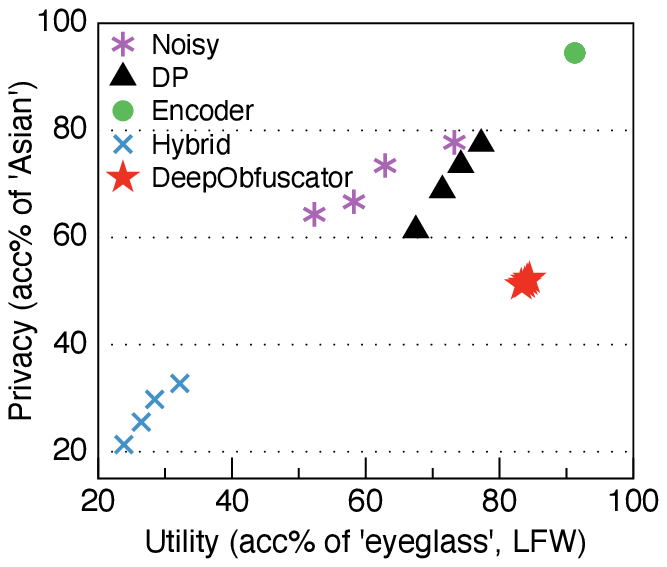}}
    \caption{Utility-privacy tradeoff comparison of DeepObfuscator with four baselines on CelebA and LFW. The higher the utility score, the better. The lower the privacy score, the better. Therefore, the best performing algorithms are those located towards the bottom right quadrant.}
    \label{fig:comparsion}
\end{figure*}

\subsection{Effectiveness of Defending Against Private Attribute Leakage}\label{subsec:eval_invisible}
\textbf{Comparison of utility-privacy tradeoff:} We compare the utility-privacy tradeoff offered by DeepObfuscator with four privacy-preserving baselines. 
In our experiments, we set `gender' and `heavy makeup' as the private attributes to protect in CelebA, and consider detecting  `smile' and `high cheekbone' as the intended classification tasks to evaluate the utility. 
With regard to LFW, we set `gender' and `Asian' as the private labels, and choose recognizing `black hair' and `eyeglass' as the intended classification tasks. 
Figure \ref{fig:comparsion} summarizes the utility-privacy tradeoff offered by four baselines and DeepObfuscator. Here we evaluate DeepObfuscator with four discrete choices of $\lambda_1\in\{1, 2, 5, 10\}$ while setting $\lambda_2=1$. Note that the evaluation metric of privacy is defined as the classification accuracy of private attributes that we aim to protect, hence, the lower accuracy of private attributes indicates a better privacy protection. Therefore, the ideal case for the utility-privacy tradeoff should be in the lower right corner, which represents achieving the maximized utility while offering the strongest privacy protection.

As Figure \ref{fig:comparsion} shows, although DeepObfuscator cannot always outperform the baselines in both utility and privacy, it still achieves the best utility-privacy tradeoff under most experiment settings. 
For example, in Figure \ref{fig:comparsion}(h), DeepObfuscator achieves the best tradeoff by setting $\lambda_1=1$. Specifically, the classification accuracy of `Asian' on LFW is 52.37\%, and the accuracy of `eyeglass' is 84.47\%. This demonstrates that DeepObfuscator can efficiently protect privacy while maintaining high accuracy of intended classification tasks. 

In other four baselines, Encoder method can maintain best utility of the extracted features, but it fails to protect privacy due to the high accuracy of private attributes achieved by the attacker. Hybrid method can provide the most effective privacy protection, but the accuracy of intended classification tasks is unacceptable.  In general, Noisy, DP and Hybrid methods offer strong privacy protection with sacrificing the utility.

\textbf{Impact of the utility-privacy budget $\lambda_1$:} An important step in the hybrid learning procedure is to determine the utility-privacy budget $\lambda_1$ and $\lambda_2$. Due to a large number of possible combinations of $\lambda_1$ and $\lambda_2$, we set $\lambda_2=1$ while varying $\lambda_1$ in this experiment. To determine the optimal $\lambda_1$, we evaluate the utility-privacy tradeoff on CelebA and LFW by setting different $\lambda_1$. Specifically, we evaluate the impact of $\lambda_1$ with four discrete choices of $\lambda\in\{1, 2, 5, 10\}$. 

For experiments using CelebA, we choose detecting `smile' and `high cheekbone' as the intended classification tasks, and `gender' and `heavy makeup' as the private attributes that the attacker aims to infer from the obfuscated features. We design 6 testing sets using different combinations of those attributes: (1) \{smile, gender\}; (2) \{high cheekbone, gender\}; (3) \{smile, high cheekbone, gender\}; (4) \{smile, gender, heavy makeup\}; (5) \{high cheekbone, gender, heavy makeup\}; (6) \{smile, high cheekbone, gender, heavy makeup\}. In fact, those six testing sets can be divided into two groups: the first three sets only contain one private attribute, and the last three sets include two private attributes. Within each group, we explore how the different numbers of the intended classification tasks will affect the protection of the private attributes. In addition, if we compare each testing set between two groups accordingly, we can investigate how the accuracy of the intended tasks will change with the number of the private attributes that need to be protected.

\begin{figure*}[t]
    \centering
    \subfigure[]{\includegraphics[scale=0.6]{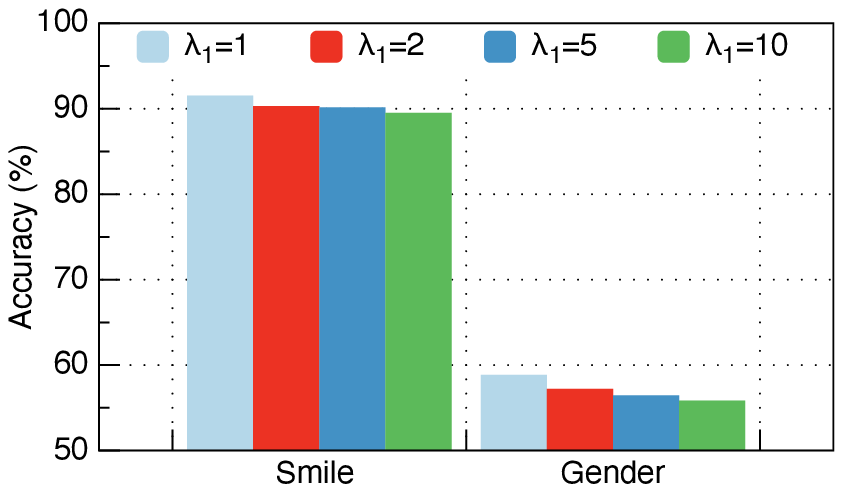}}
    \subfigure[]{\includegraphics[scale=0.6]{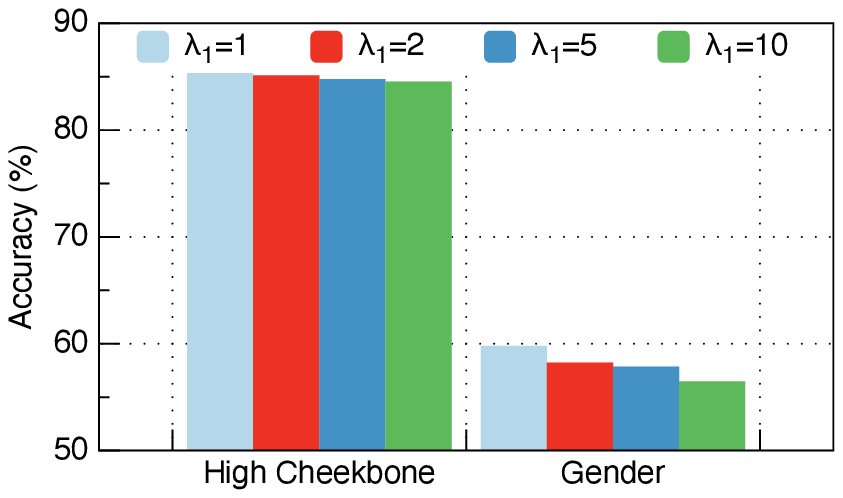}}
    \subfigure[]{\includegraphics[scale=0.6]{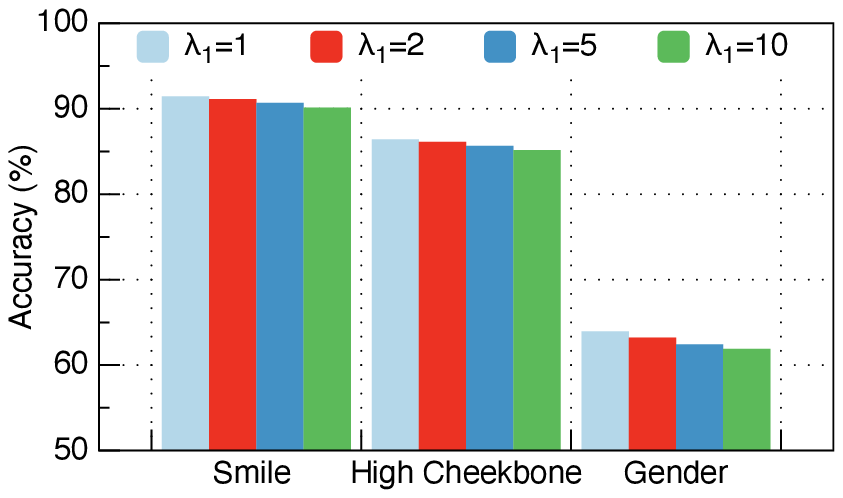}}
    \subfigure[]{\includegraphics[scale=0.6]{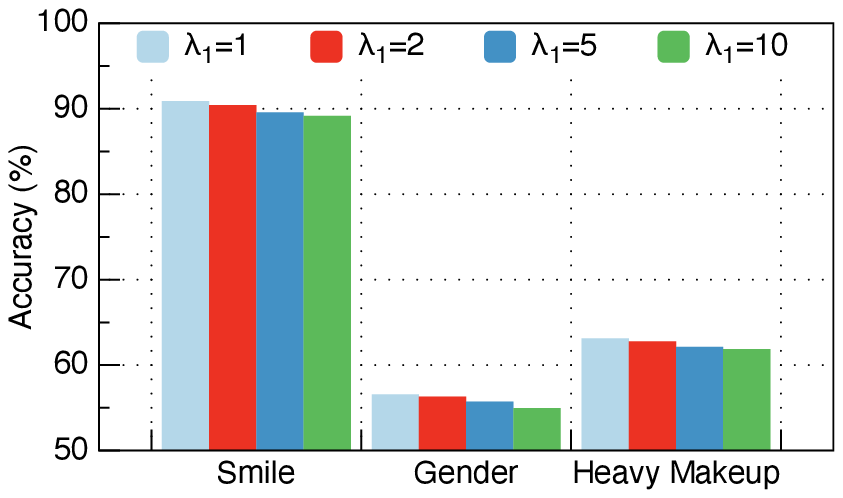}}
    \subfigure[]{\includegraphics[scale=0.6]{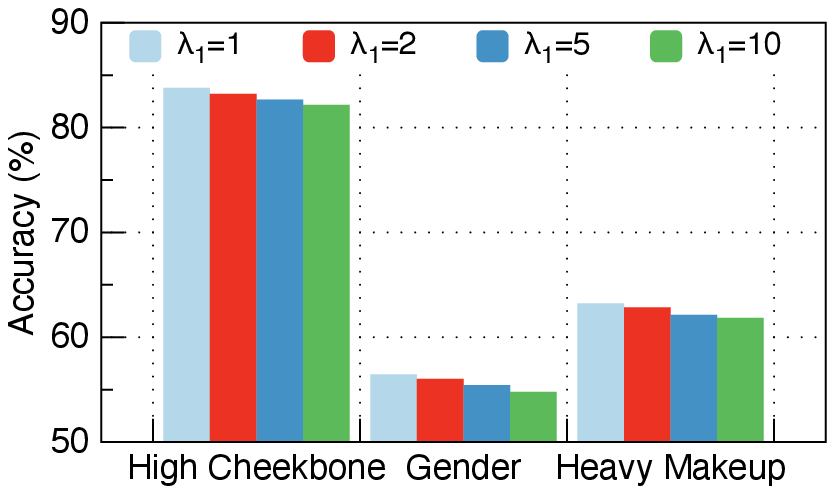}}
    \subfigure[]{\includegraphics[scale=0.6]{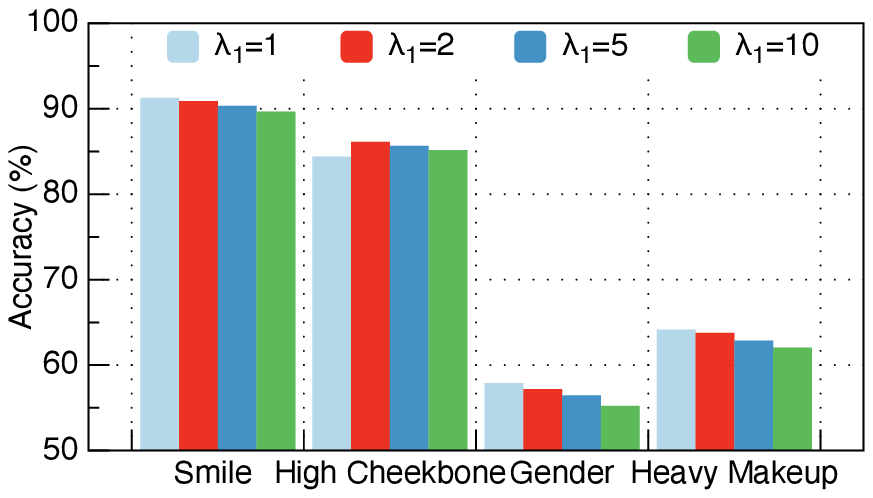}}
    \caption{The impact of the utility-privacy budget $\lambda_1$ on CelebA. ($\lambda_2=1$)}
    \label{fig:exp2}
\end{figure*}

Figure \ref{fig:exp2} shows the average accuracy of intended tasks and private attributes using the classifier and adversary classifier which are trained in the way adopted by DeepObfuscator. With the proposed adversarial training, DeepObfuscator can effectively prevent private attributes from being inferred by an attacker while only incurring a small accuracy drop on the intended classification tasks. In general, the larger $\lambda_1$ can provides a stronger privacy protection (i.e., lower accuracy of private attributes), but degrades the performance on intended classification tasks. For example, as Figure \ref{fig:exp2} shows, the accuracy of `gender' decreases from 58.85\% to 55.85\% and the accuracy of `smile' decreases from 91.53\% to 89.52\%, when increasing $\lambda_1$ from 1 to 10. However, with the increasing $\lambda_1$, the performance drop of both private attributes and intended classification tasks will be marginal.

If we compare Figure \ref{fig:exp2}(c) vs. Figure \ref{fig:exp2}(a-b) and  Figure \ref{fig:exp2}(f) vs. Figure \ref{fig:exp2}(d-e), given a particular $\lambda_1$, it can be observed that performing more intended tasks will weaken the defense against  private attribute leakage due to the intrinsic correlation between the intended tasks and the private attributes that an attacker aims to infer. Specifically, the accuracy of `gender' slightly increases from 58.85\% in Figure \ref{fig:exp2}(a) and 59.79\% in Figure \ref{fig:exp2}(b) to 63.96\% in Figure \ref{fig:exp2}(c) when setting $\lambda_1=1$. 
Similarly, the accuracy of `gender' and `heavy makeup' increase by 1\% from Figure \ref{fig:exp2}(d-e) to Figure \ref{fig:exp2}(f). In addition, by comparing Figure \ref{fig:exp2}(a-c) vs. Figure \ref{fig:exp2}(d-f), we found that protecting more private attributes leads to slight decrease in the accuracy of the intended tasks under a specific $\lambda_1$. For example, the accuracy of `smile' slightly decreases from 91.53\% in Figure \ref{fig:exp2}(a) to 90.89\% in Figure \ref{fig:exp2}(d) when setting $\lambda=1$. The reason is that the feature related to the private attributes has some intrinsic correlations to the feature related to the intended tasks. Therefore, more correlated features may be hidden if more private attributes need to be protected. As a result, the performance of the intended tasks becomes harder to maintain.

\begin{figure*}[t]
    \centering
    \subfigure[]{\includegraphics[scale=0.6]{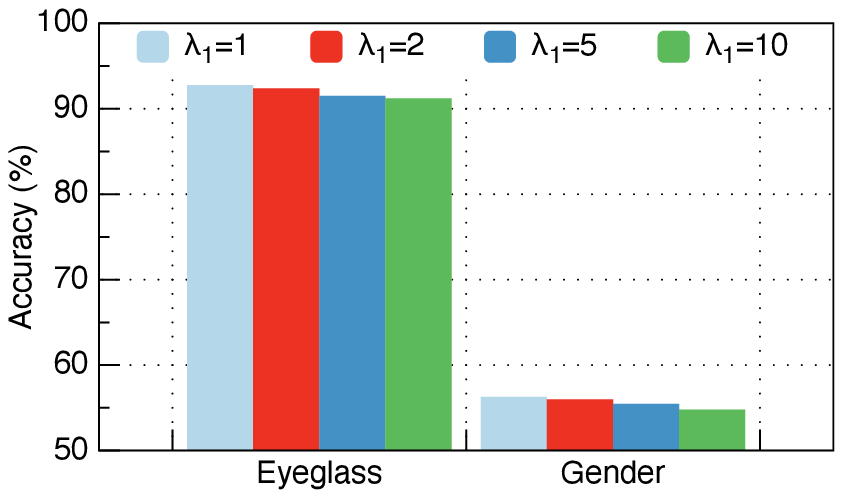}}
    \subfigure[]{\includegraphics[scale=0.6]{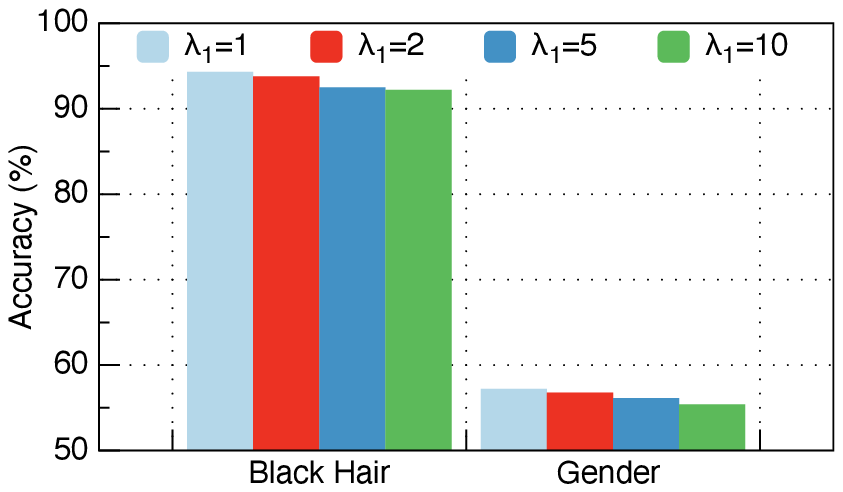}}
    \subfigure[]{\includegraphics[scale=0.6]{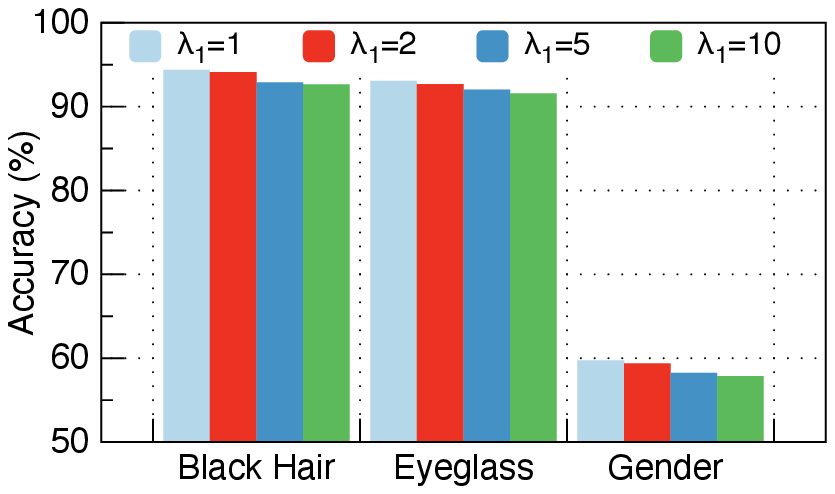}}
    \subfigure[]{\includegraphics[scale=0.6]{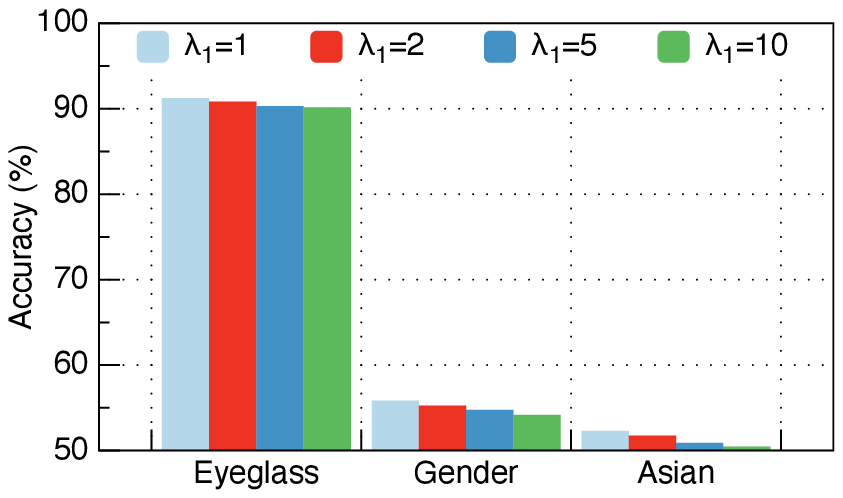}}
    \subfigure[]{\includegraphics[scale=0.6]{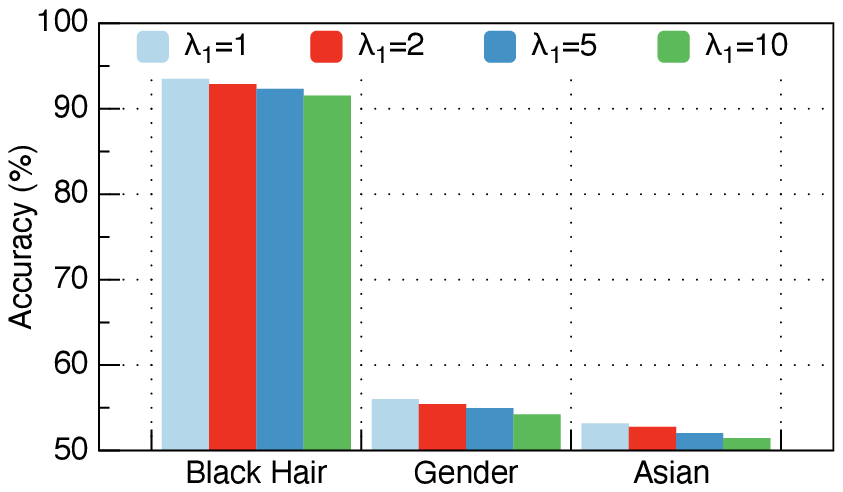}}
    \subfigure[]{\includegraphics[scale=0.6]{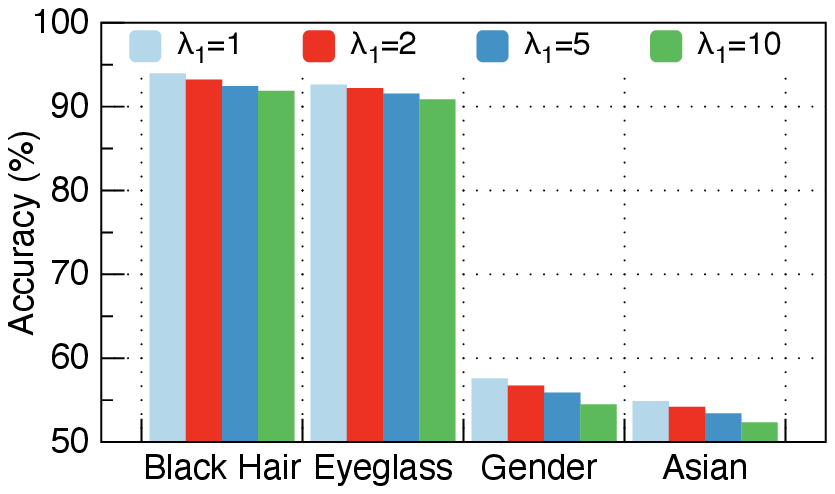}}
    \caption{The impact of the utility-privacy budget $\lambda_1$ on LFW. ($\lambda_2=1$).}
    \label{fig:exp2_lfw}
\end{figure*}

Similar to the above experiments on CelebA, in LFW, we choose recognizing `eyeglass' and `black hair' as the intended classification tasks, and `gender' and `Asian' as the private attributes that the attacker aims to infer from the obfuscated features. We also design 6 testing sets using different combinations of those attributes: (1) \{eyeglass, gender\}; (2) \{black hair, gender\}; (3) \{eyeglass, black hair, gender\}; (4) \{eyeglass, gender, Asian\}; (5) \{black hair, gender, Asian\}; (6) \{eyeglass, black hair, gender, Asian\}. As same as the settings for CelebA, those six testing sets can be divided into two groups: the first three sets only contain one private attribute, and the last three sets include two private attributes. We also conduct the same intra-group and inter-group comparisons as the above evaluations on CelebA.

Generally, we observed the result as same as that of CelebA - performing more intended tasks will weaken the defense against private attribute leakage given a specific $\lambda_1$. For example, if we compare Figure \ref{fig:exp2_lfw}(c) vs. Figure \ref{fig:exp2_lfw}(a-b),  the accuracy of `gender' slightly increases from 56.27\% in Figure \ref{fig:exp2_lfw}(a) and 57.23\% in Figure \ref{fig:exp2_lfw}(b) to 59.74\% in Figure \ref{fig:exp2_lfw}(c) with $\lambda_1=1$. Similarly, the accuracy of `gender' and `Asian' increase by 1\% from Figure \ref{fig:exp2_lfw}(d-e) to Figure \ref{fig:exp2_lfw}(f). In addition, by comparing Figure \ref{fig:exp2_lfw}(a-c) vs. Figure \ref{fig:exp2_lfw}(d-f), we also found that protecting more private attributes leads to slight decrease in the accuracy of the intended tasks. For example, the accuracy of `eyeglass' slightly decreases from 92.78\% in Figure \ref{fig:exp2_lfw}(a) to 91.25\% in Figure \ref{fig:exp2_lfw}(d) when setting $\lambda_1=1$. 

\textbf{Cross-Dataset Evaluation.} We also conduct cross-dataset evaluations by training the obfuscator using either CelebA or LFW dataset and test the performance on the other dataset. Specifically, we choose recognizing `black hair' as the intended classification task, and `gender' as the private attribute that the attacker aims to infer from the obfuscated features. As Table \ref{tab:cross} illustrates, the obfuscator that is trained using one dataset can still effectively defend against private attribute leakage on the other dataset, while maintaining the classification accuracy of the the intended classification task. For example, if we train the obfuscator using CelebA and then test it on LFW, the accuracy of `gender' decreases to 53.74\% compared with 57.23\% by directly training the obfuscator using LFW. The accuracy of `black hair' marginally increases to 94.79\% from 94.31\%. The reason is that CelebA offers a larger number of training data so that the obfuscator can be trained for a better performance. Although there is a marginal performance drop, the obfuscator that is trained using LFW still works well on CelebA. The cross-dataset evaluations demonstrate the transferability of DeepObfuscator.

\begin{table}[t]
    \centering
    \caption{Evaluate the transferability of DeepObfuscator with cross-dataset experiments.}

    \begin{tabular}{c|c|c|c}
    \hline
        \textbf{Test Dataset} & \textbf{Training Dataset} & \textbf{`gender'} & \textbf{`black hair'} \\
        \hline
        LFW & CelebA & 53.74\% & 94.79\% \\
        LFW & LFW & 57.23\% & 94.31\% \\
        \midrule
        CelebA & LFW & 59.87\% & 93.57\% \\
        CelebA & CelebA & 58.82\% & 94.88\% \\
        \hline
    \end{tabular}
    \label{tab:cross}
\end{table}

\subsection{Performance on Smartphones}
We evaluate the real-time performance of deploying the trained obfuscator on Google Pixel 2 and Pixel 3, including latency, storage and energy consumption. We randomly select 1000 images from CelebA, and feed them into the trained obfuscator which is deployed on smartphones. The averaged results are summarized in Table \ref{tb:mobile}. The learned obfuscator only occupies 5.6 MB of memory, costs 101-105 ms for extracting features, and consumes 2.7-2.8 mJ of energy for each feature extraction pass.

\begin{table}[t]
\caption{Performance of running the learned obfuscator on Google Pixel 2 and Pixel 3.}
    \centering
    \begin{tabular}{c|c|c|c}
    \hline
    \textbf{Smartphone} &  \textbf{Latency (ms)} & \textbf{Storage (MB)} & \textbf{Energy (mJ)}\\
    \hline
    Google Pixel 2    &  105 & 5.6 & 2.8 \\
    \hline
    Google Pixel 3 & 101 & 5.6 & 2.7\\
    \hline
    \end{tabular}
\label{tb:mobile}
\end{table}

\section{Case Studies}\label{sec:case}
\subsection{Case Study on Driver Behavior Recognition}\label{subsec:case}
\begin{figure*}[t]
    \centering
    \subfigure{\includegraphics[scale=0.19]{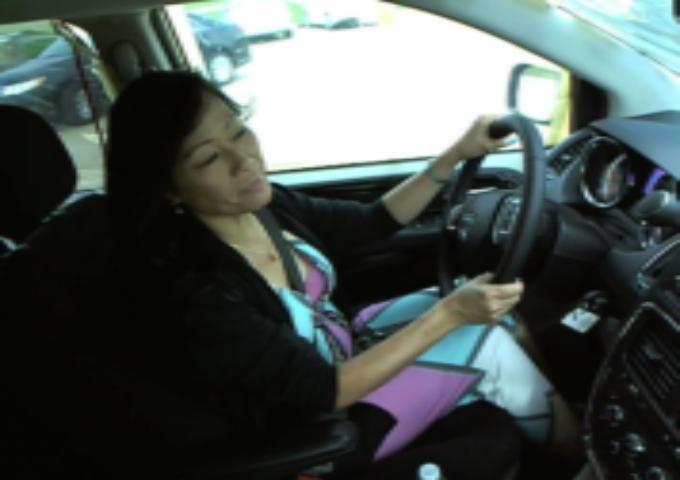}}\addtocounter{subfigure}{-1}
    \subfigure{\includegraphics[scale=0.19]{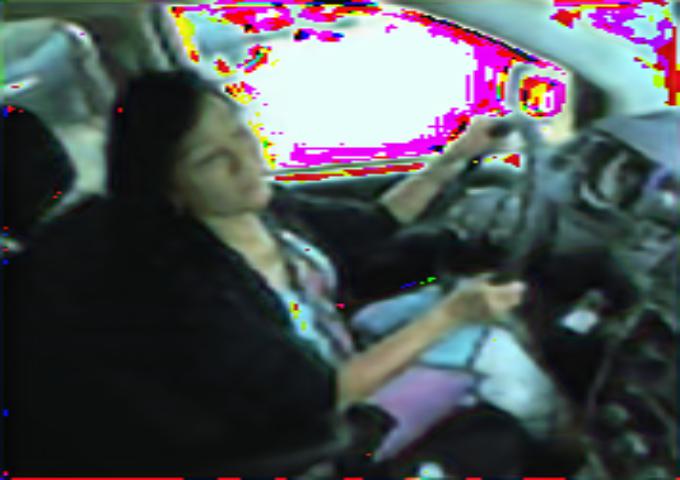}}\addtocounter{subfigure}{-1}
    \subfigure{\includegraphics[scale=0.19]{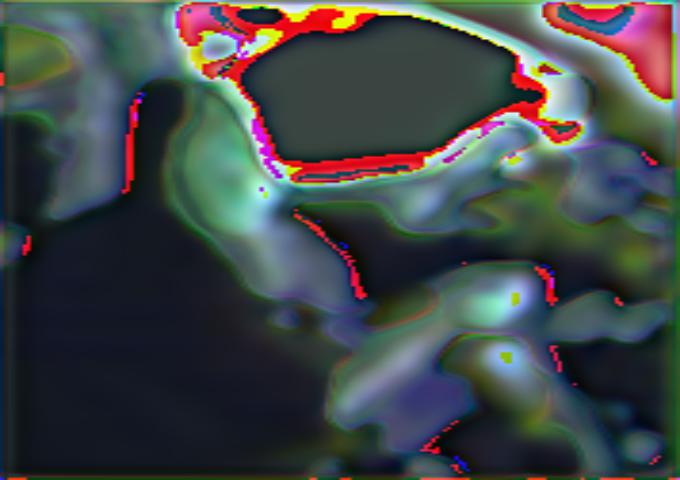}}\addtocounter{subfigure}{-1}
    \subfigure[raw images]{\includegraphics[scale=0.19]{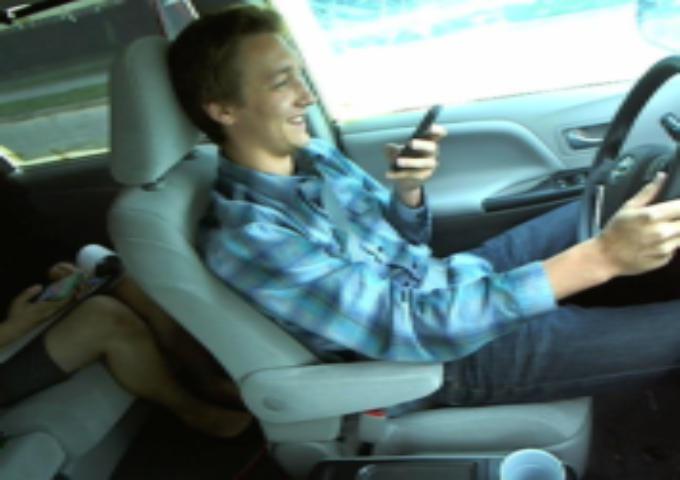}}
    \subfigure[reconstructed images with applying Encoder]{\includegraphics[scale=0.19]{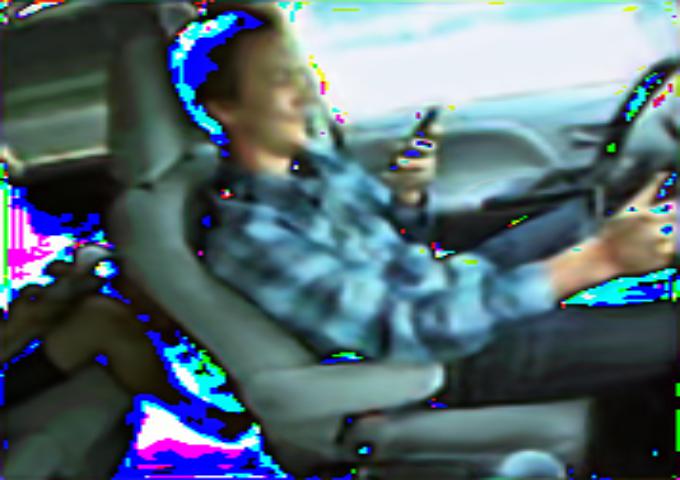}}
    \subfigure[reconstructed images with applying DeepObfuscator]{\includegraphics[scale=0.19]{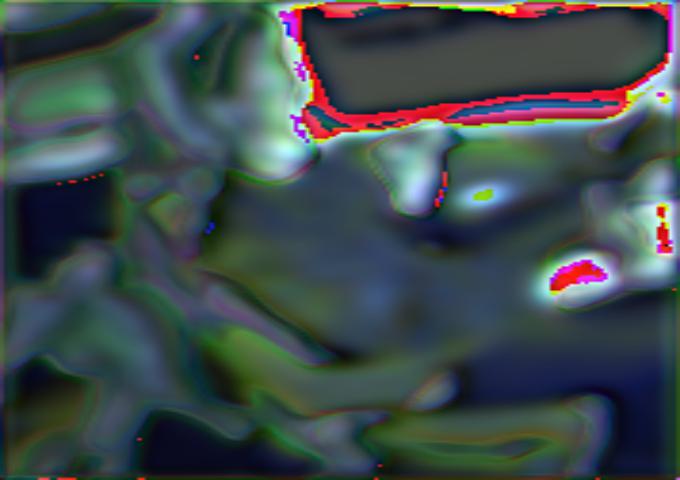}}
    \caption{Reconstructed images of the driver behavior recognition task. }
    \label{fig:driver}
\end{figure*}

 Besides facial recognition tasks which are binary classifications, performing evaluations on multi-class tasks are essential to verify DeepObfuscator’s performance. Therefore, we also evaluate DeepObfuscator on StateFarm dataset \cite{kaggle} that contains 22424 images (17939 for training, 4485 for testing) of 10 different driver behaviors from 26 people. In this experiment, driver behavior recognition is considered as an intended task while the driver's identity is a private attribute we want to protect. We consider the Encoder presented in Section \ref{sec:exp-setup} as the compared baseline. With our adversarial training method, the accuracy of driver behavior classification slightly drops  from 98.32\% to 95.49\%, but the accuracy of driver identity recognition decreases significantly from 99.97\% to 30.38\%. Figure \ref{fig:driver} shows the images reconstructed using the features encoded by DeepObfuscator, indicating that DeepObfuscator still successfully defend against the attacker's reconstruction attack.
 
\subsection{Case Study on Text Data}
DeepObfuscator can be easily extended to many other applications with various modalities of data. For example, we replace the CNN architecture with LSTM-based architecture in DeepObfuscator, and then evaluate performance on TwitterAAE dataset described in \cite{blodgett-etal-2016-demographic}. TWitterAAE consists of  166K and 10K tweets for training and testing, respectively. In our experiment, binary mention detection is considered as an intended task while the race (AAE (African-American English) or SAE (Standard American English) ) is a binary private attribute we aim to protect. The  mention detection is a binary classification task  to determine if a tweet mentions another user, i.e, classifying conversational vs. non-conversational tweets.  The architecture configurations are presented in Table \ref{tb:lstm_arch}. With applying DeepObfuscator, the mention detection accuracy marginally drops from 81.69\% to 81.38\%, however, the accuracy of race classification decreases from 79.86\% to 52.78\%, which is a totally random-guessing level.

\begin{table}[t]
\centering
\caption{The architecture configurations of LSTM-based module.}
\begin{tabular}{c|c}
\hline
\textbf{Obfuscator} & \textbf{Classifier \& Adversary Classifier}\\
\hline

Embedding-300 & 2$\times$LSTM-300\\
\cline{2-2}
LSTM-300 & FC-150\\
\cline{1-1}
& ReLU\\ 
& FC-label length\\

\hline
\end{tabular}\label{tb:lstm_arch}
\end{table}

\subsection{Comparison with PAN}
\begin{figure}[ht]
    \centering
    \subfigure{\includegraphics[scale=0.33]{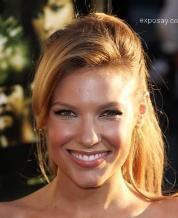}}\addtocounter{subfigure}{-1}\hspace{0.15in}
    \subfigure{\includegraphics[scale=0.33]{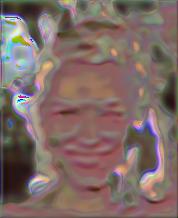}}\addtocounter{subfigure}{-1}\hspace{0.15in}
    \subfigure{\includegraphics[scale=0.33]{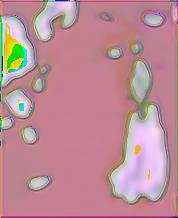}}\addtocounter{subfigure}{-1}\\
    \subfigure[raw image]{\includegraphics[scale=0.33]{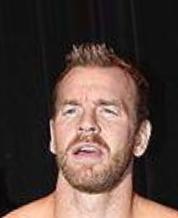}}\hspace{0.15in}
    \subfigure[reconstructed image using PAN's loss function ]{\includegraphics[scale=0.33]{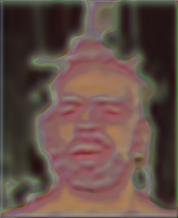}}\hspace{0.15in}
    \subfigure[reconstructed image using DeepObfuscator]{\includegraphics[scale=0.33]{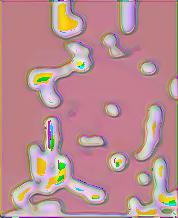}}
    \caption{Examples of reconstructed images using DeepObfuscator and PAN.}
    \label{fig:mse}
\end{figure}

PAN \cite{liu2019privacy} also provides privacy protection against the reconstruction attack by introducing a loss function for the adversary reconstructor, and the loss function is based on the Euclidean distance between a raw image and a reconstructed image. In the adversarial training, PAN aims to enlarge the difference between the raw image and the reconstructed image in terms of pixel-wise distance. In contrast to PAN, we adopt MS-SSIM as the metric to evaluate the perceptual similarity between the raw image and the reconstructed image in the loss function. More important, DeepObfuscator makes each reconstructed image similar to a crafted Gaussian noise image which is utilized in training, but significantly different from the raw image in terms of MS-SSIM. Here, we compare the effectiveness of defending against the reconstruction attack between DeepObfuscator and PAN via visualizing the reconstructed images. Specifically, we replace our loss function with PAN's Euclidean distance based loss function for comparisons, but keep anything else unchanged. As Figure \ref{fig:mse} shows, with applying the PAN's loss function, several pixels around the face are significantly changed in the reconstructed images compared with the raw images. However, the key features of the person in the reconstructed images can still be easily identified. On the contrary, it is very difficult to identify distinguishable information from the reconstructed images when applying DeepObfuscator.


\section{Discussion}\label{sec:discuss}
We evaluate the real-time performance of running the learned obfuscator on smartphones, and results show the applicability of our proposed method. However, the efficiency is a major concern about deploying the learned obfuscator on resource-constrained smartphones and edge devices. We have not perform the model optimization for the obfuscator in terms of efficiency. Numerous works have been done for model compression, such as quantization \cite{gupta2015deep}, pruning \cite{han2015deep}, etc. We propose to apply those approaches to optimize the efficiency of running the learned obfuscator on smartphone.

We evaluate DeepObfuscator on  image and text datasets, but it can be easily extended to many other applications. For example, if we replace the CNN architecture with a task-specific recurrent neural network, it is also feasible to apply DeepObfuscator to other data modalities, such as sensor data (e.g., accelerometer, gyroscope). We will evaluate DeepObfuscator using various formats of data in the future work.

Even though DeepObfuscator attains a notably better privacy-utility tradeoff than existing works, it requires the prior knowledge of primary learning tasks before training. If the primary learning tasks are changed, we need to retrain the DeepObfuscator from scratch to achieve a good privacy-utility tradeoff. However, such requirement may limit the applicability and generalization of DeepObfuscator in practice. We plan to augment DeepObfuscator with information theory-based method such that that the learned feature extractor can hide the privacy information from the intermediate representations; while maximally retaining the original information embedded in the raw data.

\section{Conclusion}\label{sec:conclusion}
We proposed an adversarial training framework DeepObfuscator for privacy-preserving image classifications by simultaneously defending against both reconstruction attack and private attribute leakage. 
DeepObfuscator consists of an obfuscator, a classifier, an adversary reconstructor and an adversary classifier. The obfuscator is trained using our proposed end-to-end adversarial training algorithm to hide sensitive information which can be exploited to reconstruct raw images and infer private attributes by an attacker. Useful features for the intended classification tasks are still retained by the obfuscator. 
The adversary reconstructor and adversary classifier play an attacker role in the adversarial training procedure, aiming to reconstruct the raw image and infer private attributes from the eavesdropped features. Evaluations on CelebA and LFW datasets show that the quality of the reconstructed images from the obfuscated features is significantly decreased from 0.9458 to 0.3175 in terms of MS-SSIM, indicating the person on the reconstructed images is hardly to be re-identified. The classification accuracy of the inferred private attributes that can be achieved by the attacker significantly drops down to a random-guessing level, but the accuracy of the intended classification tasks performed via the cloud service drops by mere 2\%. The cross-dataset evaluations demonstrate the transferability of DeepObfuscator, indicating a great practicability in the real world.

\bibliographystyle{ACM-Reference-Format}
\bibliography{reference}

\end{document}